\documentclass[print,showpacs,twocolumn,superscriptaddress]{revtex4}
\usepackage{amsmath}
\usepackage{graphicx}
\usepackage{mathrsfs}
\usepackage{amssymb}
\usepackage{amsfonts}
\usepackage{amsmath}
\usepackage{graphicx}
\usepackage{color}
\usepackage{textcomp}
\usepackage{multirow}

\newcommand{\ybodd}{$^{171}$Yb }

\newcommand{\yboddion}{$^{171}$Yb$^+$ }

\newcommand{\YbOdd}{${}^{171}$Yb${}^+$}

\begin{document}
\title{Quantum gates using electronic and nuclear spins of Yb$^{+}$ in a magnetic
field gradient}
\author{Kunling Wang}
\affiliation{State Key Laboratory of Magnetic Resonance and Atomic
and Molecular Physics, \\ Wuhan Institute of Physics and
Mathematics, Chinese Academy of Sciences, and Wuhan National
Laboratory for Optoelectronics, Wuhan 430071, China}
\affiliation{Graduate School of the Chinese Academy of Sciences,
Beijing 100049, China}
\author{Michael Johanning}
\affiliation{Fachbereich Physik, Universit{\"a}t Siegen, 57068,
Siegen, Germany}
\author{Mang Feng}
\email{mangfeng@wipm.ac.cn} \affiliation{State Key Laboratory of
Magnetic Resonance and Atomic and Molecular Physics, \\ Wuhan
Institute of Physics and Mathematics, Chinese Academy of Sciences,
and Wuhan National Laboratory for Optoelectronics, Wuhan 430071,
China}
\author{Florian Mintert}
\affiliation{Freiburg Institute for Advanced Studies,
Albert-Ludwigs-Universit\"at Freiburg, Albertstraße 19, 79104
Freiburg, Germany}
\author{Christof Wunderlich}
\email{wunderlich@physik.uni-siegen.de}
\affiliation{Fachbereich Physik, Universit{\"a}t Siegen, 57068,
Siegen, Germany}

\pacs{03.67.Lx, 42.50.Dv}

\begin{abstract}
An efficient scheme is proposed to carry out gate operations on an
array of trapped Yb$^+$ ions, based on a previous proposal using
both electronic and nuclear degrees of freedom in a magnetic field
gradient. For this purpose we consider the Paschen-Back regime
(strong magnetic field) and employ a high-field approximation in
this treatment. We show the possibility to suppress the unwanted
coupling between the electron spins by appropriately swapping states
between electronic and nuclear spins. The feasibility of generating
the required high magnetic field is discussed.
\end{abstract}

\maketitle

\section{Introduction}

Quantum computing (QC) based on nuclear spins has attracted
considerable interest over the past decades due to the long
coherence time of nuclear spins. Since they only weakly interact
with their environment, nuclear spins are well suited for storing
quantum information, and, for the same reason, difficult to
manipulate. As a result, to carry out QC, one may work with
ensembles of nuclear spins \cite {nmr} or employ a hyperfine
interaction to manipulate individual nuclear spins by electron spins
\cite {lukin1}. The former, using mature techniques of nuclear
magnetic resonance, has become a test bed for models and schemes of
QC and quantum simulations \cite{Peng2009,Du2010}, while the
exploitation of the latter is still at an early stage.

Furthermore, there have been proposals combining nuclear and
electronic spins in solid-state systems, such as doped silicon
substrates \cite{kane} and doped fullerenes \cite{wolfgang}. To
distinguish nuclear and electronic degrees of freedom, one has to
introduce a strong magnetic field and work in the Paschen-Back
regime.  An experimental demonstration of the manipulation of a
nuclear spin ensemble with about 10,000 ions in a Penning trap under
magnetic field 0.8 T was reported by Bollinger et al. \cite{rf}. The
manipulation of individual nuclear spins had only been achieved in
diamond nitrogen-vacancy centers, with qubits encoded in the
$^{13}$C or $^{15}$N nuclear spins near the electron spin \cite
{jele1}.

Here, we propose a scheme, using trapped $^{171}$Yb$^{+}$ ions, to
encode qubits in both electronic and nuclear spins of trapped atomic
ions for QC and quantum simulations, following a previous idea
\cite{ca}. Different from already accomplished experimental work
with trapped ions \cite{Blatt2008}, we consider quantum logic
operations on ions in the Paschen-Back regime with qubits encoded in
nuclear spins $I$ and auxiliary qubits in electron spins $S=1/2$.
This combines the long decoherence time of nuclear spins with
efficient manipulation and readout using electron spins. Quantum
information is stored in nuclear spins and is only swapped into
electronic spins for single-qubit gates and conditional quantum
dynamics with two and more ions. Thus, quantum information remains
well protected from ambient noise fields that otherwise would give
rise to decoherence.

Swapping quantum information between nuclear and electron spins is
accomplished using microwave radiation. Subsequent conditional
quantum dynamics between electron spins and individual addressing of
electron spins may be done using laser light \cite{cz,ms}, or, in
the presence of a spatially varying magnetic field, using again
microwave radiation
\cite{Mintert2001,Wunderlich2002,Wunderlich2003,jason,Johanning2009,Ospelkaus2008,Wang2009}.
A magnetic field gradient induces spin-spin coupling
\cite{Wunderlich2002,Wunderlich2003,jason} between electronic spins
that can be used for quantum logic gates, and, in addition allows
for ions to be addressed in frequency space. The latter approach is
useful, since it avoids technical and fundamental difficulties when
using laser light for coherent operations
\cite{Mintert2001,Wunderlich2002,Ozeri2007,Ospelkaus2008}. In
addition, it allows for conditional quantum dynamics without
stringent requirements on the cooling of the ions' vibrational
motion \cite{Loewen2004}.

The key point of our proposal is the suppression of unwanted
coupling between electronic spins in the magnetic field gradient by
swapping quantum information between nuclear and electron spins by
microwave radiation. As a result, the overhead operations, such as
refocusing pulses, in \cite{ca} is no longer necessary. In what
follows, we first investigate the use of nuclear and electron spins
of atomic ions with and without the high-field approximation, from
which we know how well the quantum gate is performed in the real
experimental situation. Also, a spatially varying magnetic field is
included. Then we show how two-qubit gating is achieved without the
need for compensating for unwanted couplings between electron spins.
Furthermore, a detailed discussion of how the required magnetic
field and the magnetic gradient can be achieved is given.

For concreteness, the present work considers as an example the use
of \YbOdd{} ions for quantum information
science\cite{Huesmann1999,Hannemann2002,Wunderlich2003,Balzer2006,Kielpinski2006,Maunz2007,Cetina2007}.
$^{171}$Yb$^{+}$ features a nuclear spin of one half and thus
provides the simplest hyperfine structure with several potential
qubits where the experimenter can choose both magnetically sensitive
and insensitive qubit transitions (to first order). The hyperfine
qubits can be directly manipulated by a resonant microwave field or
by using an optical Raman transitions.  Recent interest in trapping
Yb$^{+}$ is, to some extent, motivated also by the fact that the
experimenter benefits from low priced diode lasers for
photoionization \cite{Johanning2010} and for all transitions
relevant to Doppler cooling, qubit initialization and state
selective detection(See Fig. 1).

\begin{figure}[t]
\centering
\includegraphics[width=1.0\columnwidth]{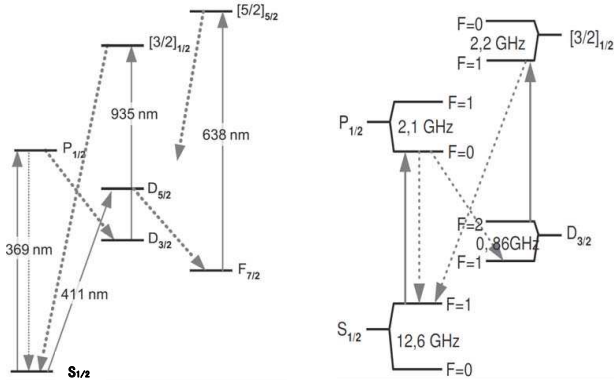}
\caption{Energy-level schemes of $^{172}$Yb$^{+}$ (Left) and
$^{171}$Yb$^{+}$ (Right), where the energy gaps are not drawn to scale.}
\label{fig:yb172}
\end{figure}

The paper is structured in four sections. In section II we present
the Hamiltonian describing a linear Coulomb crystal of ions and
justify the high-field approximation. In section III we will
demonstrate efficient QC operations using both nuclear and
electron spins, the experimental feasibility of which is discussed
in detail in section IV. We give a brief summary in the last
section.

\section{The system and the Hamiltonian}

We consider an array of trapped ions in a linear trap in the
presence of a magnetic field gradient, whose Hamiltonian in units of
$\hbar=1$ is written as,
\begin{eqnarray}
\label{eq:H} H = \sum_i \Omega_S^i S_z^i + \sum_i \Omega_I^i I_z^i +
A \sum_i (S_x^i I_x^i + S_y^i I_y^i + S_z^i I_z^i)    \nonumber \\
-\frac{1}{2} \sum_{i<j} J_{ij} \hat{S}_z^i \hat{S}_z^j,
~~~~~~~~~~~~~~~~~~~~~~~~~~~~~~~~~~~~~~~~~~~~~~
\end{eqnarray}
where $S_{k}$ and $I_{k}$ ($k=x, y, z$) are, respectively, the spin
operators of the electron spin ($S=1/2$) and the nuclear spin $I$.
For \YbOdd{} we have $I=1/2$, the hyperfine coupling constant is $A=
12.645$ GHz and the Larmor frequencies are given by
$\Omega_{S}=g_{S}\mu_{B}B=28 B$ GHz and $\Omega_{I}=g_{n}\mu_{B}B =
-7.5 B$ MHz with $B$ the strength of the magnetic field in Tesla
experienced by the ion. $J_{ij}$ is the coupling between the
electron spins of the trapped ions under the magnetic field gradient
\cite{Wunderlich2002,Wunderlich2003,jason}. We leave out the
vibrational modes here, because additional radiation fields applied
to swap information between nuclear and electron spins only drives
carrier transitions, that is, these fields do not couple vibrational
and spin states. The nuclear spin couplings are also neglected as
they are very small compared to other terms.

The magnetic field is applied along the trap axis, so the $i$th ion
experiences the magnetic field with
$$\vec{B}=[B_{0} + bz]\hat{e}_{z},$$
with $B_{0}$ the strength of the magnetic field at the origin,
$b=\partial{B}/\partial{z}$  the magnetic field gradient, and
$\hat{e}_{z}$ the unit vector along the trap axis. For the spin-spin
coupling we have
\begin{equation}
\label{eq:J} J_{ij} = \sum_{l=1}^{N} \frac{2}{m \nu_l^2}
D_{il} D_{jl} \frac{\partial \Omega^i_S}{\partial z} \frac{\partial
\Omega_S^j}{\partial z},
\end{equation}
where $m$ is the mass of trapped ions, \(\nu_l^2 = \nu_z^2 \mu_l\)
with $\nu_z$ the axial frequency of the trap and $\mu_l$ the
eigenvalue of potential Hessian matrix. $D$ is the unitary
transformation matrix that diagnonalizes the Hessian matrix and
$\Omega_S^i$ depends on magnetic gradient $b$. With respect to the
original expression of $J_{ij}$ in \cite {Wunderlich2002,
Wunderlich2003}, Eq. (2) seems formally larger by 4 times, which is
because we use angular momentum operators here instead of the Pauli
operators.

For our purpose, we first consider the single-ion case to justify the
high-field treatment. In the Paschen-Back regime, the Hamiltonian of a
single ion is obtained by reducing Eq. (1),
\begin{equation}
H_{0}=\Omega_{S}S_{z}+\Omega_{I}I_{z}+ A(S_{x}I_{x}+S_{y}I_{y}+S_{z}I_{z}). \label{e.1}
\end{equation}
Assuming a magnetic field $B=1$ T, we plot the energy-level
structure determined by Eq. (3) in Fig. 2 for \YbOdd, where the
eigenstates include some superpositions due to the x- and y-terms of
the hyperfine interaction. In contrast, the conventional treatment,
to simplify the problem, is the exclusion of the x- and y-terms of
the hyperfine interaction under the high-field approximation, i.e.,
\begin{equation}
H_{1}=\Omega_{S}S_{z}+\Omega_{I}I_{z}+ AS_{z}I_{z}. \label{e.2}
\end{equation}
Since each term in this Hamiltonian is diagonal, the eigenstates of the Hamiltonian
are ones of $S_{z}$ or $I_{z}$.

\begin{figure}[htb]
\begin{center}
\includegraphics[width= 0.98\columnwidth]{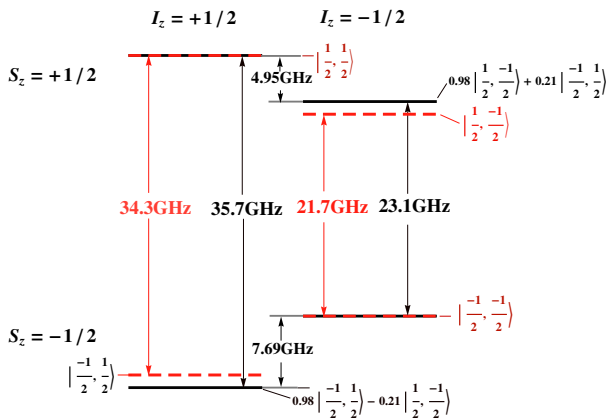}
\end{center}
\caption{(Color online)Angular momentum states in the ground state of a
$^{171}$Yb$^{+}$ ion in a magnetic field $B=$1 T (i.e. in the
Paschen-Back regime). The levels drawn with the black solid lines
and the red dashed lines are, respectively, from the
exact Hamiltonian Eq. (3) and the approximated Hamiltonian Eq. (4).}
\label{fig1}
\end{figure}

In order to carry out single-qubit gates and conditional quantum
gates with nuclear spins (that are used as a quantum memory), it is
necessary to transfer the nuclear spin's state to the electron spin
and vice versa. Since the SWAP gate could be performed by
appropriate CNOT gate sequences, e.g., SWAP =
CNOT$_{IS}$CNOT$_{SI}$CNOT$_{IS}$ =
CNOT$_{SI}$CNOT$_{IS}$CNOT$_{SI}$ \cite{ca} with CNOT$_{ab}$
implying the control $a$ and target $b$, we consider below the
necessary CNOT gates, which could be accomplished by radiating the
ion with appropriate $\pi$ pulses \cite{ca}. The key point is the
consideration of the level shifts due to hyperfine interaction.

Taking a CNOT$_{SI}$ gate as an example, under the high-field
approximation with the magnetic field 1 T, we may radiate the ion by
a 6.31 GHz microwave pulse, yielding the flip between $|1/2,
1/2\rangle$ and $|1/2, -1/2\rangle$ (See energy levels in red by
dashed lines in Fig. \ref{fig1}). This pulse does not lead to
transitions between the levels $|-1/2, 1/2\rangle$ and $|-1/2,
-1/2\rangle$ due to detuning. As a result, the flip of the nuclear
spin is controlled by the electronic spin. The idea is easily
extended to perform a CNOT$_{IS}$ with the nuclear spin as the
control qubit. If the exact treatment is used instead of the
high-field approximation, however, the involvement of x- and y-terms
of hyperfine interaction in Eq. (\ref{e.1}) makes the CNOT gates
considered above less perfect.

As shown by the black solid lines in Fig. \ref{fig1}, to achieve the
transition $|1/2, 1/2\rangle \leftrightarrow |1/2, -1/2\rangle$, we
may employ the pulse with frequency 4.95 GHz, which actually
leads to
$$|1/2, 1/2\rangle_{ex} \leftrightarrow 0.9776|1/2, -1/2\rangle_{ex}
+ 0.2103|-1/2, 1/2\rangle_{ex},$$ where $|\dots\rangle_{ex}$ means
the state under exact evolution. So the ion would leak to the
unwanted state $|-1/2, 1/2\rangle$ state with probability 0.04.

To have a preferable quantum gate in a realistic experiment under the
high-field approximation, we have to first justify the condition for
Eq. (4). To this end, we require the level splittings under the
high-field approximation to be identical to the exact situation,
which makes the theoretical treatment closer to the realistic
operation. So we introduce effective gyromagnetic ratios to replace
the natural gyromagnetic ratios in the treatment of Eq. (4). By
numerics, we have found the effective Larmor frequencies should be,
$$\Omega_S'\approx \gamma_S' B_0 ~~\mathrm{GHz}$$
and
$$\Omega_I'\approx -\gamma_I' B_0 ~~\mathrm{GHz}$$
for 1 T $< B_0 < 5$ T. Effective gyromagnetic ratios read
$\gamma_S'\approx (28.1+5.5 e^{-1.5 B_0/B_1})$ and $\gamma_I'\approx
-(0.085+5.5 e^{-1.5 B_0/B_1})$ (with $B_{1}= 1$ T). With these
effective Larmor frequencies, Eq. (4) becomes
\begin{equation}
H_{1}=\Omega'_{S}S_{z}+\Omega'_{I}I_{z}+ AS_{z}I_{z}. \label{e.3}
\end{equation}
Eq.({\ref{e.3}}) is a good approximation to Eq. (3) which gives
nearly the same energy levels. In Fig. 3 we show a highter fidelity 
is achieved for a CNOT gate with increasing magnetic field.

\begin{figure}[htb]
\begin{center}
\includegraphics[width= 0.8\columnwidth]{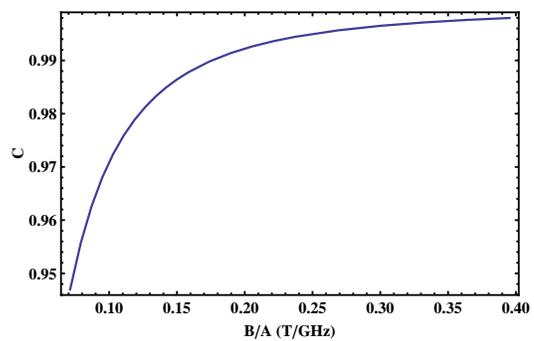}
\end{center}
\caption{(Color online) Fidelity C of the CNOT$_{SI}$ gating with
respect to the magnetic field strength $B$ divided by hyperfine
coupling constant A. Here $C=|\langle\Psi|\Psi_{exact}\rangle|^{2}$
with $|\Psi\rangle$ and $|\Psi_{exact}\rangle$ the evolved
wavefunctions under Eq. (\ref{e.3}) with effective Larmor
frequencies and under Eq. (\ref{e.1}). This result with scaled
magnetic field can be applied to other ions with I=1/2 and S=1/2 in
a good approximation (i.e., neglecting  the difference between
nuclear Larmor frequencies of different ions). For \ybodd$^+$, the
magnetic field changes within the range (0.9 T, 5 T).} \label{fig2}
\end{figure}

From now on we use the approximate Hamiltonian Eq. (\ref{e.2}) with
effective Larmor frequencies to simplify the treatment in multi-ion
situation. We have to emphasize that the purpose of the
approximation we employ is, on the one hand, to keep consistent with
the conventional treatments in previous works \cite{wolfgang,ca}, on
the other hand, to have a clear physical picture for our gate
operations. Since the nuclear and electronic spins are never
completely decoupled in real case, we justified the approximation in
the above treatment to try to find a trade-off for achieving
accurate and coherent gate operations.

\section {Quantum gating using S-I swap}

Consider a string of trapped ions in the presence of a spatially
varying magnetic field along the z-direction, as described by Eq.
(1). In the Paschen-Back regime, we may neglect the x- and y-terms
in the hyperfine interaction. So the Hamiltonian is reduced to
\begin{equation}
\label{eq:H2} H_2 = \sum_i \Omega_S^{'i} S_z^i + \sum_i
\Omega_I^{'i} I_z^i + A \sum_i S_z^i I_z^i-\frac{1}{2} \sum_{i<j}
J_{ij} S_z^i S_z^j.
\end{equation}
The nuclear spins, due to negligible coupling with each other,
remain the same as in the single ion case. But we have to pay more
attention to the electron spins, which are coupled due to the
magnetic field gradient. Because of these $J$ couplings, the
transition frequency of a given ion depends on the electron spin
states of others. So with an increasing number of ions the spectrum
of the ion chain becomes more complex.

A good candidate system for gate operations should have the coupling
between qubits well controlled. In the absence of a magnetic field
gradient, trapped ionic qubits interact by coupling to the common
vibrational modes mediated by suitably tuned laser light
\cite{cz,ms,Blatt2008}. Here, in the presence of a magnetic field
gradient, the electron spins' coupling, that reaches well beyond
nearest neighbors, is to be used for conditional quantum dynamics.
Other QC proposals that make use of nuclear and electron spins
usually assume only nearest-neighbor coupling \cite{wolfgang}. For
our present trapped ion model, however, the interactions between the
ions are significantly beyond the nearest-neighbor couplings.

Previously proposed solutions to this problem include: (1)
refocusing operations applied simultaneously with the gating pulses
on the trapped ions \cite {ca} and (2) additional potentials applied
on the trapped ion \cite {jason,Wunderlich_H2009}. The former
solution is based on exact knowledge of the undesired coupling, and
overhead in this method increases quickly with the number of qubits.
The latter requires micro-structured electrodes traps to
shape the effective potential confining the ions. In order to
produce sizeable $J$ couplings in the system, the electrodes' axial
extension should be of the order of 10 micrometers or smaller and
the distance between the electrodes' surface and the ions needs to
be of similar magnitude.

In what follows we present an alternative approach to accomplish
high-fidelity two-qubit gating making use of the two spins available
in each trapped ion. Since both the electron spin and the nuclear
spin are initially polarized, the quantum information during the QC
implementation is only stored in one of them and the other remains
well polarized. When an ion is active, i.e., operated for gating or
readout, the quantum information is encoded in the corresponding
electron spin and the corresponding nuclear spin remains well
polarized. When the ion turns passive, the quantum information is
swapped to the nuclear spin for storage, and the electronic spin
becomes well polarized.

\begin{figure}[htb]
\begin{center}
\includegraphics[width= 0.98\columnwidth]{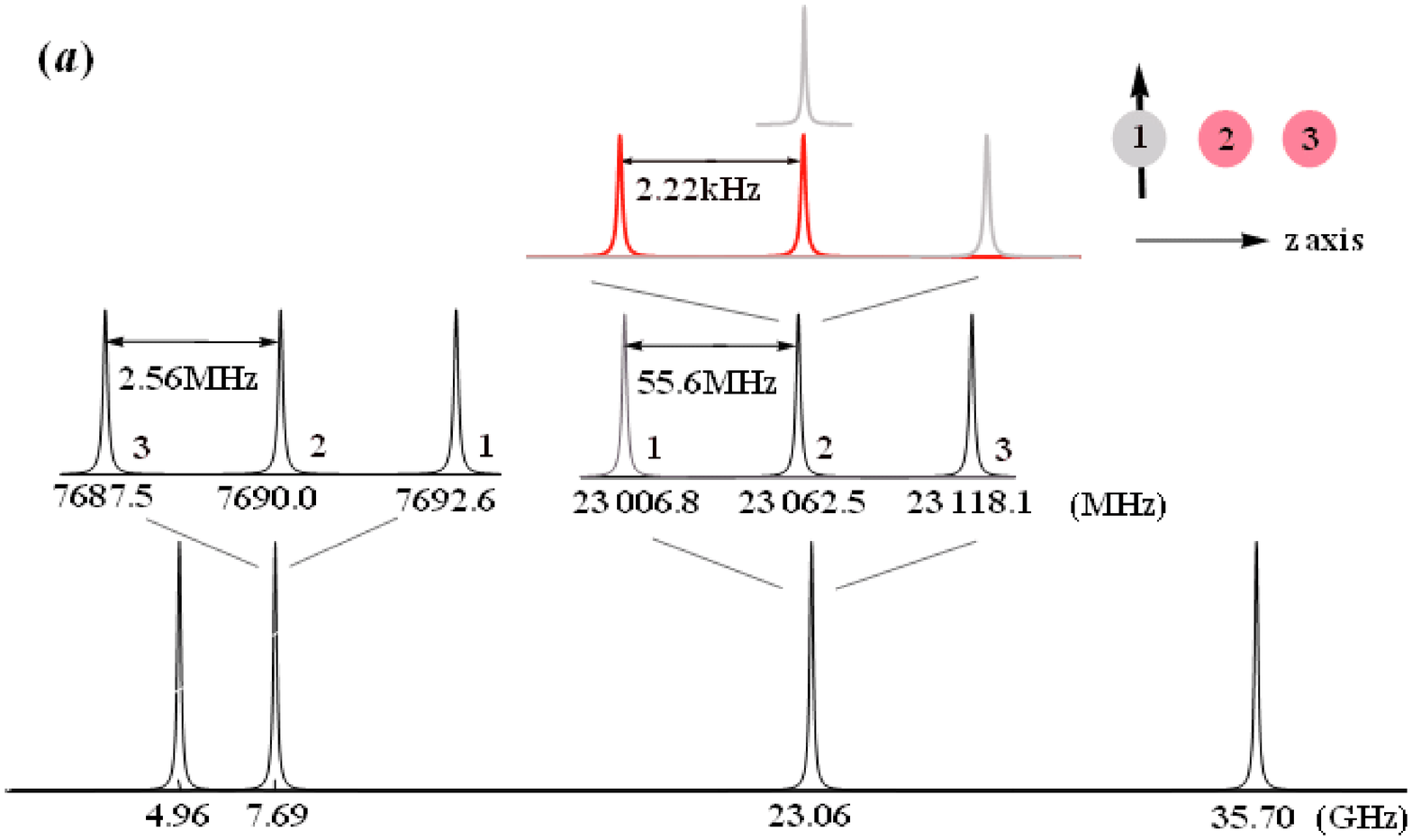}
\includegraphics[width= 0.98\columnwidth]{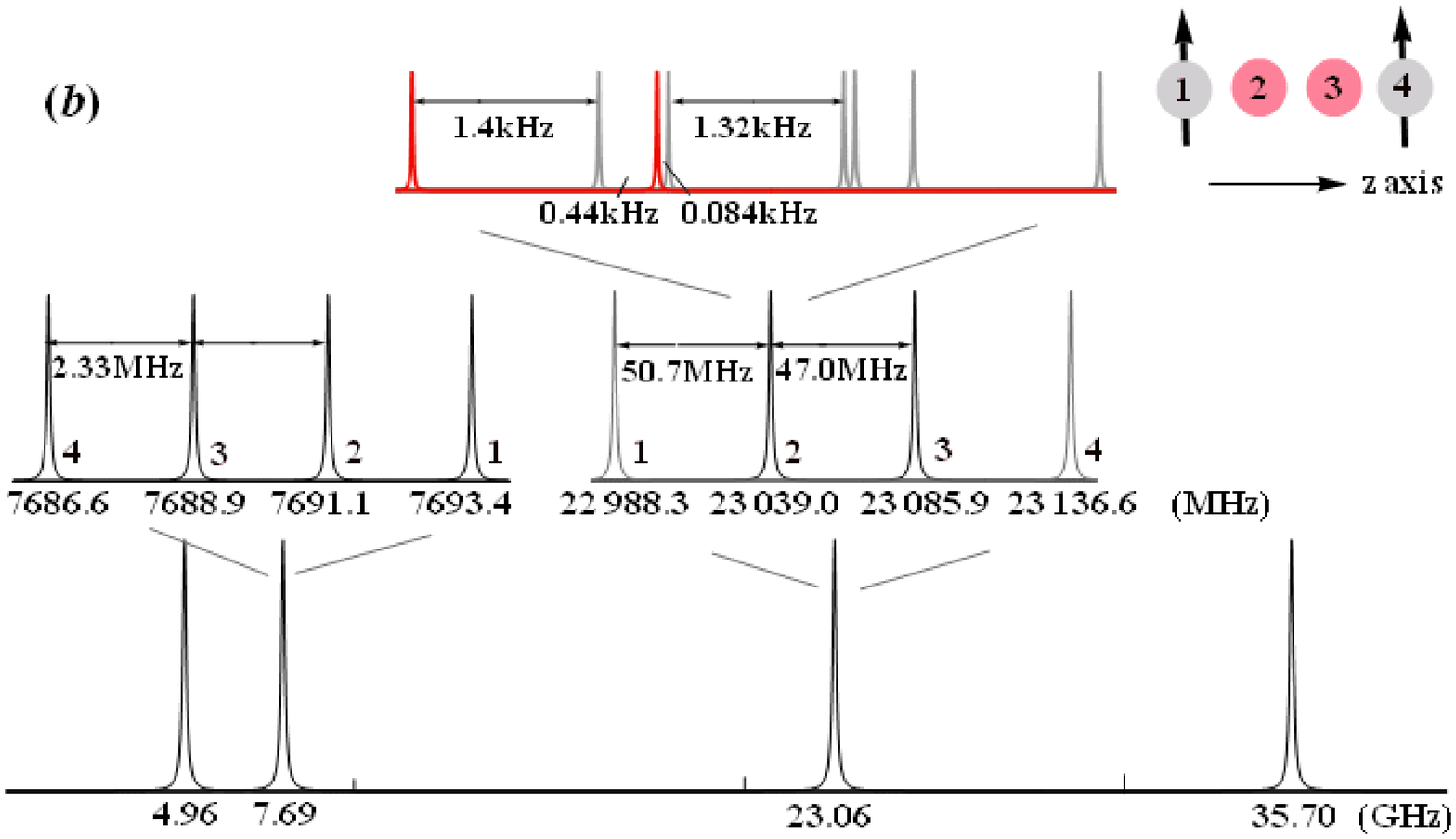}
\end{center}
\caption{(Color online) Spectrum of trapped ions  with magnetic field $B_0$=1 T,
field gradient b=500 T/m, and axial trap frequency $\nu_z$=600 kHz,
where (a) is for three ions and (b) for four ions. The lowest trace
extends over all resonances; the middle trace shows the splitting
due to a magnetic field gradient, and the upper trace depicts the
splitting due to $J$ coupling between the ions. Resonant frequencies
are marked in lowest and middle traces, and splittings are labeled
in the middle and upper traces (Some splittings with no labeling due
to symmetry). The ions (Nos. 2 and 3) in red stand for active ions
and others in gray for passive ions. The frequency lines in red are
the ones left after the passive electronic spins are polarized.}
\label{fig4}
\end{figure}

Consider two active ions in an array of ions. Since the electronic
spins of all other ions are well polarized and their nuclear spins,
encoding the quantum states, have no interaction with other ions,
the two active electronic spins only experience a frequency shift by
the polarized electronic spins of the rest passive ions. Compared to
the case with the passive electron spins in arbitrary superposition,
this scheme makes the spectra much simpler. As a result, quantum
gating would be much easier since refocusing unwanted interactions
or operations with locally shaped electrostatic potentials in
micro-traps are no longer necessary. In addition, quantum states
stored in nuclear spin degrees of freedom are more robust to
decoherence than in electronic counterpart, which helps to store
quantum information in higher fidelity.

Fig.{\ref{fig4}} demonstrates the cases with three- and four-ions as
examples, where two of them are active and the rest are passive. By
polarizing the passive electron spins, there are only two frequency
lines left for the active spins, which are shifted with respect to
the original positions by corresponding $J$ couplings.

We have simulated the two-qubit gating for three and four
$^{171}$Yb$^{+}$ trapped in a line under a magnetic field gradient,
as shown in Table I. The gating time is dependent on the magnetic
field gradient, but not on the magnetic field itself. A stronger
magnetic field is required just for a better gate manipulation.

\begin{table}[htb]
\caption{ Two-qubit $\mathrm{CNOT_{S_1S_2}}$ gating time $T$ for
three and four trapped $^{171}$Yb$^{+}$ in a strong magnetic field
$B_{0}=1$ T with axial trap frequencies $\nu_{z}=600$ kHz and 200 kHz for
different magnetic field gradients $b$ and distances $\Delta
z_{\min}$. $J$ represents the nearest-neighbor coupling for $J_{ij}$
in Eq. (6) and data for 4 ions are regarding the middle two ions. N
stands for number of ions.}
\begin{center}
\begin{tabular}[c] {|l|l|l|l|l|l|} \hline
$\nu_z$(kHz) & N & $b$ $($T/m$)$ & $\Delta z_{\min}$ $(\mu$m$)$ & $J$ $($kHz$)$ & $T$ $($ms$)$ \\
\hline
\multirow{6}{*}{600} & $ 3$ & $50$ & $4.15$ & $0.0444$ & $70.8$\\
                     & $ 3$ & $100$ & $4.15$ & $0.178$ &$17.7$\\
                     & $ 3$ & $300$ & $4.15$ & $1.60$ & $1.97$\\
                     \cline{2-6}
                     & $ 4$ & $50$ & $3.50$ & $0.0368$ & $85.2$\\
                     & $ 4$ & $100$ & $3.50$ & $0.147$ & $21.3$\\
                     & $ 4$ & $300$ & $3.50$ & $1.33$ & $2.37$\\
\hline
\multirow{6}{*}{200} & $ 3$ & $50$ & $8.63$ & $0.399$ & $7.87$\\
                     & $ 3$ & $100$ & $8.63$ & $1.60$ &$1.97$\\
                     & $ 3$ & $300$ & $8.63$ & $14.38$ & $0.218$\\
                     \cline{2-6}
                     & $ 4$ & $50$ & $7.28$ & $0.332$ & $9.47$\\
                     & $ 4$ & $100$ & $7.28$ & $1.33$ & $2.37$\\
                     & $ 4$ & $300$ & $7.28$ & $11.94$ & $0.263$\\
\hline
\end{tabular}\end{center}
\label{tab.gs}
\end{table}

\section{Experimental feasibility}

In order to implement QC and quantum simulations with nuclear spins
of trapped ions as described above, a strong and highly stable
magnetic field is required. If conditional quantum dynamics is
carried out based on magnetic gradient induced coupling
\cite{Mintert2001,Wunderlich2002,Wunderlich2003,Johanning2009,Ospelkaus2008}
using microwave or radio frequency radiation, instead of laser
light, then the applied magnetic field in addition needs to vary
spatially. In addition, a magnetic field gradient allows for
individual ion addressing in frequency space using rf or microwave
radiation \cite{Johanning2009,Wang2009}. Below, we will discuss the
feasibility of creating the required strong field using Yb$^{+}$
ions as a concrete example.

Qubits may be encoded in the simple hyperfine structure of the
isotope \yboddion. Gradients can be achieved by using permanent
magnets, for example, in quadrupole configuration or by using a pair
of anti-Helmholtz coils or shaped planar current geometries known
from magnetic traps for neutral atoms \cite{Folman2002,Fortagh2007}.
The field noise can be reduced by superconductive materials
\cite{Wang2009}.

In micro-structured traps (two- and three-dimensional), the required
magnetic gradient extends only over a limited volume, and thus does
not necessarily require strong fields. The optimization of the field
geometry using current carrying micro-structures to reach a maximum
gradient with a limited current (or dissipated power) appears not
too different from the task of optimizing for maximum field. To
reach magnetic fields in the Tesla range, however, massive cooling
of the current carrying structures would be necessary.

A three-dimensional ion trap has been designed that allows for
generating gradients of up to 100 T/m \cite{Wunderlich_H2009}.
Details on this trap will be given elsewhere. For neutral atom
trapping, two-dimensional current structures were exploited to
create flexible magnetic field configurations and magnetic gradients
\cite{Groth,Folman2002,Fortagh2007}. A good thermal contact between
current carrying structures and the substrate allows to efficiently
remove any thermal intake due to ohmic heating, resulting in
enormous possible current densities of $j_{\rm{max}} \approx
10^{11}$~A/m$^2$ \cite{Groth} and allows for versatile and fast
switchable fields and gradients. With appropriate cooling of the
substrate, we expect gradients in the range of 100-300~T/m to be
possible for two-dimensional traps that are currently under
development in our laboratory at Siegen.

An alternative straightforward solution for producing {\em both}
strong magnetic fields {\em and} high magnetic gradients would be a
pointed yoke which, in its proximity, would create a combination of
both. It is desirable, however, to create homogeneous and
inhomogeneous parts independently: the gradient is necessary for the
addressing of single ions and the coupling between ionic qubits, but
it impedes the efficient cooling of the ion chain as a whole, since
the microwave transition, which is also required during cooling to
avoid optical pumping would be different for each ion. It can thus
be advantageous to switch the gradient on during manipulation only,
and set it to zero during cooling, and potentially during read-out.
The strong offset field, however, that would indiscriminately shift
all resonance frequencies over widely spread frequency bands upon
switching, remains on at all times. Therefore, in what follows, we
will focus on an independent creation of homogeneous and
inhomogeneous magnetic fields.

\subsection{Possible approaches to creation of high magnetic fields}

Methods to generate strong magnetic fields suitable for ion trap QC
include:
\begin{itemize}
\item Permanent magnets produce stable and low noise fields, which are simple
and inexpensive, but cannot be switched on and off nor be tuned
directly - the field on a given point however, could be changed by
changing the position or orientation of a magnet.
\item Current carrying structures on the other hand can produce time dependent fields,
but require high power current supplies, and in most cases
stabilization and cooling.
\item Superconducting current carriers have a better inherent stability
but require a high initial experimental effort and costs for setting
up a cryostat, and thereby in the long term exhibit high operating
costs.

When using type-II super-conductors \cite{Wilson}, the persistence
of the magnetic field in superconducting coils is usually viewed as
an advantage and reduces ac field noise, which, in terms of
coherence time is, of course, desirable. On the other hand,
permanent magnets offer intrinsic low field noise and the
persistence of superconducting magnets can become an obstacle, in
case one intends to modulate the magnetic field periodically using
additional current carrying structures. This can be useful, for
instance, in order to insert temporal phases with a homogeneous
magnetic field and thus homogeneous or global cooling. Slow
variations are possible, but not on the time-scale of the typical
repetition rate (order of 100 Hz) of data taking.

\end{itemize}

For a static and homogeneous offset field, we will focus on
permanent magnets, where much progress  has been made during the
last decades \cite{new15}: Not only did the maximum remanence of
commercially available permanent magnets increase substantially (the
remanence of Nd$_{\rm{2}}$Fe$_{\rm{14}}$B can reach values of up to
$B_r=1.22$~T \cite{hcp}), but also progress was made in the task of
maximizing the field with a given magnet material by choosing a
suitable mounting geometry. Our investigations concerned with the
creation of strong magnetic fields, therefore, focus on the usage of
optimized magnet arrangements to exceed the surface flux of a single
magnet. This can be achieved for example by pole and yoke design or
in Halbach arrangements \cite{HighMagFields,Halbach1980}.

\subsection{The Halbach structure}

Ideally, a Halbach structure consists of an infinitely long
magnetized cylinder of continuously varying magnetization direction
with inner diameter $r_i$ and outer diameter $r_o$. The
magnetization of a Halbach dipole points along the angle $2\phi$ for
an infinitesimal cylinder segment at angle $\phi$. This results in a
cancelation of fields outside of the cylinder and a homogeneous
magnetic field inside the cylinder with the magnitude
\begin{equation}
B = B_r \ln\left(\frac{r_o}{r_i}\right), \label{eq:halbach-analytic}
\end{equation}
with $B_r$ being the remanence.

A conclusion for the generation of high magnetic fields can be seen
from this idealized analytic expression
Eq.~(\ref{eq:halbach-analytic}: for limited outer dimensions, a
smaller inner diameter allows for larger field strength. The trap
structure and vacuum housing inside the Halbach cylinder cannot be
made arbitrarily small, effectively limiting the achievable field
strength.

\begin{figure}[t]
\centering
\includegraphics[width= 0.85\columnwidth]{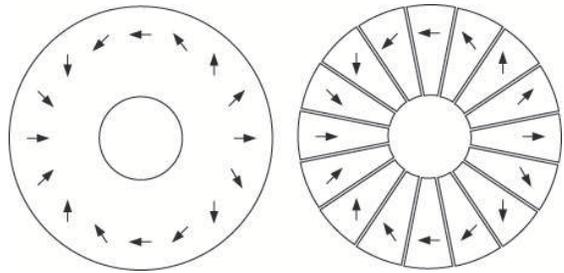}
\caption{Schematics for the ideal Halbach configuration (left) and a
segmented approximation with $N=16$ segments (right), which
theoretically delivers more than 97\% of magnetic field strength in
the inner cylinder as compared to the ideal structure. Black arrows
indicate the local direction of the magnetic field.}
\label{fig:halbach}
\end{figure}

This structure can be replaced, for the ease of fabrication, by a
segmented cylinder, made of $N$ homogeneously magnetized segments,
as shown in Fig. \ref{fig:halbach}, where the field inside becomes
\begin{equation}
B = B_r \left(\frac{\sin(2\pi/N)}{2\pi/N}\right)
\ln\left(\frac{r_o}{r_i}\right), \label{eq:halbach-segments}
\end{equation}
and with $N=16$ segments one
can reach already 97\% of the field strength of the idealized case.

The finite length of any real Halbach structure, too, contributes to
a reduction of the $B$ field according to
\begin{equation}
B = B_r \ln\left(\frac{r_o}{r_i}\right) - B_r f(z_0),
\label{eq:halbach-finite}
\end{equation}
where the reduction factor $f(z_0)$ depends on the length $z_0$ and
the radii $r_i$ and $r_o$ of the cylinder as
\[
f(z_0) = \left[ \frac{z_0}{2\sqrt{z_0^2 r_o^2}} -
\frac{z_0}{2\sqrt{z_0^2 r_i^2}}+ \ln\left(\frac{z_0 + \sqrt{z_0^2
r_o^2}}{z_0 + \sqrt{z_0^2 r_i^2}}\right)\right]
\].

\begin{figure}[t]
\centering
\includegraphics[width= 0.95\columnwidth]{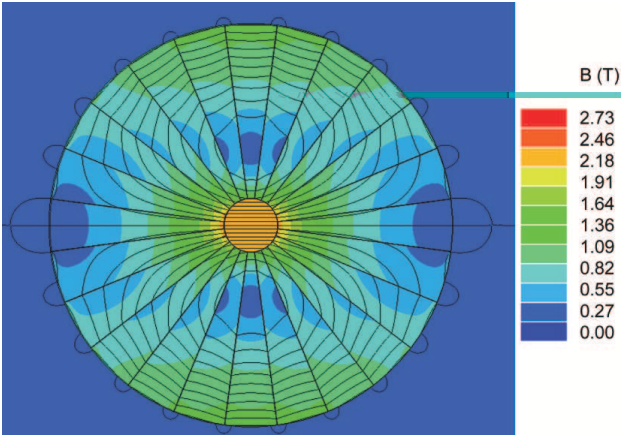}
\caption{(Color online) Numerical simulation with a Halbach dipole ring with an
inner diameter of 5~cm and an outer diameter of 50~cm, yielding a
homogeneous offset field of 2.3~T, in contrast to the analytical
model, yielding above 2.8~T. In addition, a set of flux lines is
shown.} \label{fig:halbachBCombined}
\end{figure}

We carried out numerical simulations for a structure with 16
segments with a cylindrical inner volume of a diameter 5~cm using
NdFe35, with a  retentivity of $B_r=1.23$~T. A very homogeneous
field of 2.3~T is obtained when the outer diameter is limited to
50~cm (see Fig. \ref{fig:halbachBCombined}).

Even stronger fields can be achieved in three-dimensional structures
which follow the  same concept, namely Halbach spheres, at the
expense of constrained optical access to the high-field region
\cite{RareEarthMagnets}. The theoretical field for such an
arrangement is given by
\begin{equation}
B = \frac{4}{3} B_r \ln\left(\frac{r_o}{r_i}\right),
\label{eq:halbach-sphere}
\end{equation}
which is already larger by a factor $1/3$ than the field created by
a comparable cylinder, but the low optical access makes this choice
less attractive, unless the whole detection is placed inside the
sphere.

In such structures, the field can exceed the maximum coercivity of
the permanent magnet and locally reverse its magnetization (if
aligned unfavorably), thus imposing another practical limit on
attainable field strengths. This can be avoided, by replacing parts
of the magnets with materials with high coercivity (and often lower
remanence), and in this way, magnetic fields exceeding 4.5~T have
been created \cite{Cugat1998}.

The maximum attainable field strength  is limited by material
properties as remanence and coercivity, which are usually functions
of temperature. For example for NdFeB, the temperature coefficient
of the remanence is -0.1~\%~K$^{-1}$, and the temperature
coefficient for the coercivity is -0.6~\%~K$^{-1}$, allowing for
substantial improvements even if cooling only to liquid nitrogen
temperatures \cite{Sagawa1984, Tanaka2006}.

\subsection{Effective Potential with magnetic field}

Given the substantial strength of magnetic fields considered here,
its impact on the ion's motion is a priory not necessarily
negligible. To assess the impact of a strong magnetic field, we
consider here the dynamics of an ion with mass $m$ and charge $e$ in
the presence of both an rf trap potential and a strong magnetic
field. An ion's momentum $\vec{p}$ is thus replaced by
$\vec{p}-\frac{e}{m}\vec{A}$ with the vector potential $\vec{A}$
satisfying $\vec{B}=\nabla\times\vec{A}$. The exact solution of the
equations of motion contains quickly oscillating terms associated
with the ion's micro-motion at the frequency $\Omega_t$ of the rf
trap drive. Averaging over the micro-motion will yield solutions
characterized by an effective (pseudo-) potential
\cite{Ghosh,Werth}.

The Hamiltonian reads ${{\cal H}={\cal H}_{0}+{\cal
H}_{1}\cos\Omega_{t} t}$, with
\begin{equation}\begin{array}{cccl}
\vspace{1em} &
{\cal H}_{0} & = & \frac{(\vec{p}-e\vec{A})^{2}}{2m}+\frac{1}{8}m\Omega_{t}^{2}a(x^{2}+y^{2}) \\
\mbox{and}\hspace{2em} & {\cal H}_{1} & = &
-\frac{1}{4}m\Omega_{t}^{2}q(x^{2}-y^{2})\ ,
\end{array}\end{equation}
where $a$ and $q$ are the usual stability parameters characterizing
the trapping potential \cite{Ghosh}.

The corresponding Schr\"odinger equation can be solved with the
ansatz
\begin{equation}
\Psi(x,y,t)=\Phi(x,y,t)e^{-i\alpha(t)V_{t}(x,y)}
\label{ansatz}\end{equation} of a slowly varying wave function
$\Phi(x,y,t)$ and the quickly oscillating phase
$\alpha=\frac{1}{\hbar\Omega_{t}}\sin\Omega_{t} t$. Averaging the
time-dependent Schr\"odinger equation over the interval ${\delta
t=2\pi\Omega_{t}^{-1}}$ and taking $\Phi$ to be constant over this
period yields
\begin{equation}
i\hbar\frac{\partial\Phi}{\partial t}={\cal H}\Phi\ ,
\end{equation}
with
\begin{equation}
{\cal
H}=\frac{1}{2m}(p_{x}^{2}+p_{y}^{2})+\frac{1}{2}m\omega_{r}^{2}(x^{2}+y^{2})+\frac{1}{2}\omega_{c}(p_{x}y-p_{y}x)
\label{hamiltonmitbfeld}\end{equation} where the following averages
have been used:
$\frac{\Omega_{t}}{2\pi}\int_{0}^{2\pi\Omega_{t}^{-1}} \alpha dt=0$,
$\frac{\Omega_{t}}{2\pi}\int_{0}^{2\pi\Omega_{t}^{-1}} \alpha^{2} dt=\frac{1}{2\hbar^{2}\Omega_{t}^{2}}$
and $\nabla{\cal H}_{1}=-\frac{1}{2}m\Omega_{t}^{2}q\left[{x\atop
-y}\right]$.

Introducing the standard creation and annihilation operators
\begin{equation*}\begin{array}{cccc}
&
a_{k}^{\dagger} & = & \frac{1}{2\hbar}\left(\sqrt{m\omega_{r}}k-\frac{i}{\sqrt{m\omega_{r}}}p_{k}\right)\\
\mbox{and}\hspace{2em} & a_{k}           & = &
\frac{1}{2\hbar}\left(\sqrt{m\omega_{r}}k+\frac{i}{\sqrt{m\omega_{r}}}p_{k}\right)
\end{array}
\hspace{3em}k=x, y
\end{equation*}
yields
\begin{equation}
{\cal
H}=\hbar\omega_{r}(a_{x}^{\dagger}a_{x}+a_{y}^{\dagger}a_{y}+1)+
\frac{i}{2}\hbar\omega_{c}(a_{x}^{\dagger}a_{y}-a_{y}^{\dagger}a_{x})\
,
\end{equation}
{\it i.e.} a Hamiltonian for which $x$ and $y$ components are
coupled with $\omega_c=eB/m$. This coupling can easily be resolved
by introducing new creation and annihilation operators
\begin{equation}
a_{+}=\frac{1}{\sqrt{2}}(a_{x}+ia_{y})
\hspace{1em}\mbox{und}\hspace{1em}
a_{-}=\frac{1}{\sqrt{2}}(a_{x}-ia_{y})\ ,
\end{equation}
in terms of which the Hamiltonian reads
\begin{equation}
{\cal H}=
\hbar(\omega_{r}+\frac{1}{2}\omega_{c})(a_{+}^{\dagger}a_{+}+\frac{1}{2})+
\hbar(\omega_{r}-\frac{1}{2}\omega_{c})(a_{-}^{\dagger}a_{-}+\frac{1}{2})\
.
\end{equation}
That is, analogously to the classical case, there are two decoupled
modes with shifted frequency $\omega_{r}\pm\frac{1}{2}\omega_{c}$,
and the motion in $z$-direction is unaffected by the magnetic field.

\section{Discussion and conclusion}

In summary, we have proposed to encode quantum information in
nuclear spins of trapped atomic ions and considered the feasibility
of nuclear spin quantum information processing using trapped
$^{171}$Yb$^{+}$ ions in a linear ion trap. Employing both nuclear and
electron spins provides not only the combination of robust storage
of quantum information with efficient quantum gating, but also a
good way to suppress the undesired coupling between electron spins.
The discussion of possible methods to generate the required magnetic
field indicates that this scheme is feasible with currently or
near-future available ion-trap techniques.

This scheme could also be applied to other candidate ions, such as
$^{43}$Ca$^{+}$ \cite {Blatt2008,ca} or $^{9}$Be \cite{Blatt2008}.
Since the hyperfine coupling in those ions is much smaller than in
$^{171}$Yb$^{+}$, it is possible to satisfy the high-field
approximation using a lower magnetic field. On the other hand,
because the nuclear spins of those ions are not 1/2 (i.e.
$^{43}$Ca$^{+}$ with $I=$7/2 and $^{9}$Be with $I=$3/2), it would be
more complicated to encode and manipulate qubits in nuclear spins.
For example, as studied in \cite{ca}, the nuclear spin flipping
operation would be more complex and take a relatively longer time.
Particularly, when x and y terms of the hyperfine coupling are
considered in the Hamiltonian, a very high magnetic field is
required to obtain the desired fidelity. Using the similar
calculation for the $\mathrm{CONT_{IS}}$ gate for 7/2 nuclear spin
$^{43}$Ca$^{+}$ in \cite{ca}, we have found that 5T magnetic field
are necessary to get an effective operation as good as that in the
present paper for $^{171}$Yb$^{+}$. For $^{9}$Be$^{+}$, at least 1 T
magnetic field is needed to reach a good high-field approximation.

In addition, without the magnetic field gradient, our scheme would
still work using laser light for the electronic spin operations
using the Cirac-Zoller model \cite {cz} or M{\o}lmer-S{\o}rensen
model \cite {ms}. Due to involvement of nuclear spins, more qubits
could be employed in the system with the same numbers of ions
trapped compared to previous schemes using only electron spins.

\section*{Acknowledgements}

This work is supported by National Natural Science Foundation of
China under Grant No. 10774163 and No. 10974225, by the National
Fundamental Research Program of China under Grant No 2006CB921203,
by the Deutsche Forschungsgemeinschaft, the EU STREP PICC, and
secunet AG.

\end{document}